\documentclass[a4paper,twoside]{article}

\usepackage{authblk}
\usepackage{graphicx}
\usepackage{booktabs}
\usepackage{amsmath}
\usepackage{amssymb}
\usepackage{rotating}
\usepackage{textcomp}
\usepackage{xcolor}
\usepackage[numbers]{natbib}
\usepackage[margin=2cm]{geometry}

\title{Conductivity influence on interfacial waves in liquid metal
  batteries and related two-layer systems}

\author[1]{T.~Weier}
\author[2]{I.~Grants}
\author[1]{G.M.~Horstmann}
\author[1]{S.~Landgraf}
\author[1]{M.~Nimtz}
\author[1]{P.~Personnettaz}
\author[1]{F.~Stefani}
\author[1]{N.~Weber}

\date{January 28, 2020}

\affil[1]{Helmholtz-Zentrum Dresden - Rossendorf, Bautzner Landstr. 400, 01328 Dresden, Germany}
\affil[2]{Institute of Physics, University of Latvia, Salaspils-1, LV-2169, Latvia}

\begin{document}
\maketitle

\begin{abstract}
  Fluid flows in liquid metal batteries can be generated by a number
  of effects. We start with a short overview of different driving
  mechanisms and then address questions specific to the metal pad role
  instabilities in three-layer systems. We focus on the role of the
  conductivity distribution in the cell, noting at the same time that
  interfacial tension should be considered as well for smaller
  cells. Following this discussion, numerical results on the
  excitation of interfacial waves in two-layer liquid metal systems
  with miscibility gaps bearing an interface normal electric current
  are presented. Confirming recent results from the literature, we
  find that magnetic damping plays a decisive role for strong vertical
  magnetic fields. In addition, boundary conditions for the electric
  field strongly influence critical currents and growth rates.
\end{abstract}

\section*{Introduction}
Liquid metal batteries (LMBs) are possible candidates for meeting the
\begin{figure}[b]
\includegraphics[width=\textwidth]{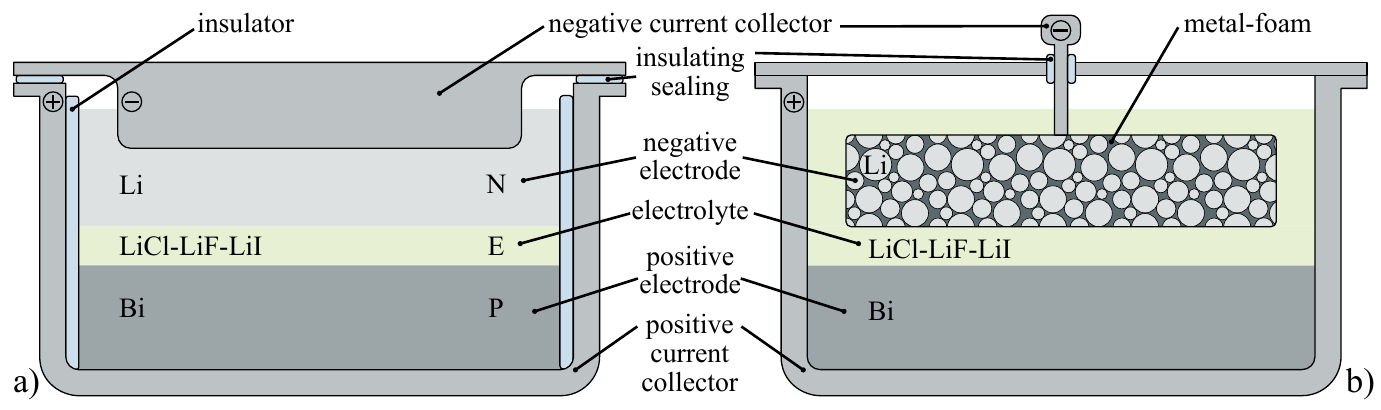}
\caption{Sketch of liquid metal batteries implemented as a
  differential density cell (a) and a cell featuring a retainer
  (metal-foam in this case) for the negative electrode (b).}
\label{fig:LMB_sketch}
\end{figure}
large-scale storage demands of future electricity systems relying
mainly on volatile sources as wind power and photovoltaics. Single LMB
cells contain three liquid layers in stable density stratification:
typically an alkaline or earth alkaline metal as the negative
electrode (N) on the top, a molten salt mixture as the electrolyte (E)
in the middle and a heavy post-transition metal or metalloid at the
bottom (P), see Fig.~\ref{fig:LMB_sketch}a. A comprehensive overview on
LMBs including their history is provided by Kim et
al.~\cite{KimBoysenNewhouseEtAl:2013}.

Due to the molten active materials and the fused electrolyte,
fluid mechanics plays a much larger role for LMBs than for most other
battery types. Current densities routinely approach
1\,A/cm\textsuperscript{2} and can be as high as
13\,A/cm\textsuperscript{2}. The accompanying Joule heating is
typically concentrated in the electrolyte layer. For many material
combinations used in LMBs, the electrolyte's electrical conductivity
is about four orders of magnitude lower than those of the electrodes,
but see Fig.~\ref{fig:conductivities}a for exceptions. This heat
source in the middle of the cell will generally provide good mixing in
the electrolyte and - depending on the height ratios - strong
convection in the negative electrode
\cite{ShenZikanov:2015}. Temperature gradients in the positive
electrode should be stabilizing in most cases.

Current flow in batteries is accompanied and influenced by mass
transport. This phenomenon is particularly pronounced in the positive
electrode and is either generating or damping convection, depending on
the current direction, see
\cite{PersonnettazLandgrafNimtzWeberWeier:2019a,PersonnettazLandgrafNimtzWeberWeier:2020} for
examples. An estimation of the thermal and compositional Rayleigh numbers
will typically predict a dominating solutal convection.

The high current density might as well give rise to electro-vortex
flows as discussed in more detail in
\cite{LiuLiStefaniWeberWeier:2019}. Their action could be beneficial,
e.g., improving mixing in the positive electrode, but care should be
taken to prevent disruptions of the electrolyte layer and to avoid
resulting short circuits.

\begin{figure}[b!]
  \setlength{\unitlength}{\textwidth}
  \begin{picture}(1.0,0.42)
    \put(0.0,0.07){%
      \includegraphics[width=0.39\textwidth]{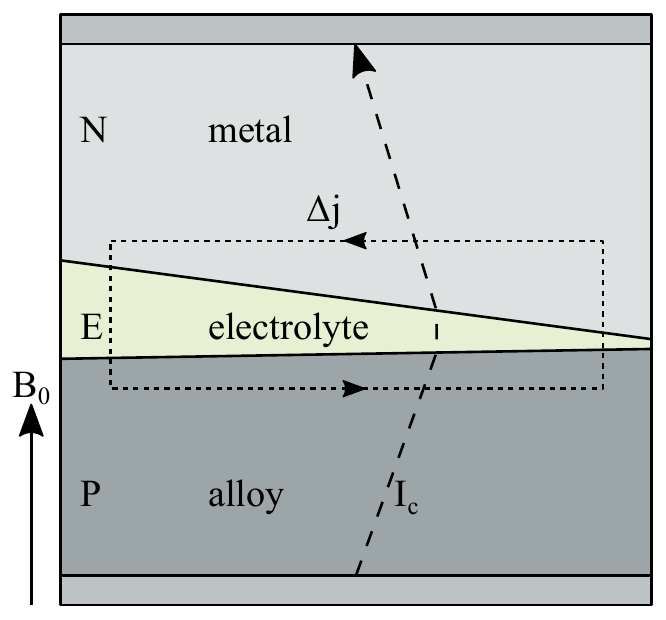}
    }
    \put(0.4,0.0){%
      \includegraphics[width=0.6\textwidth]{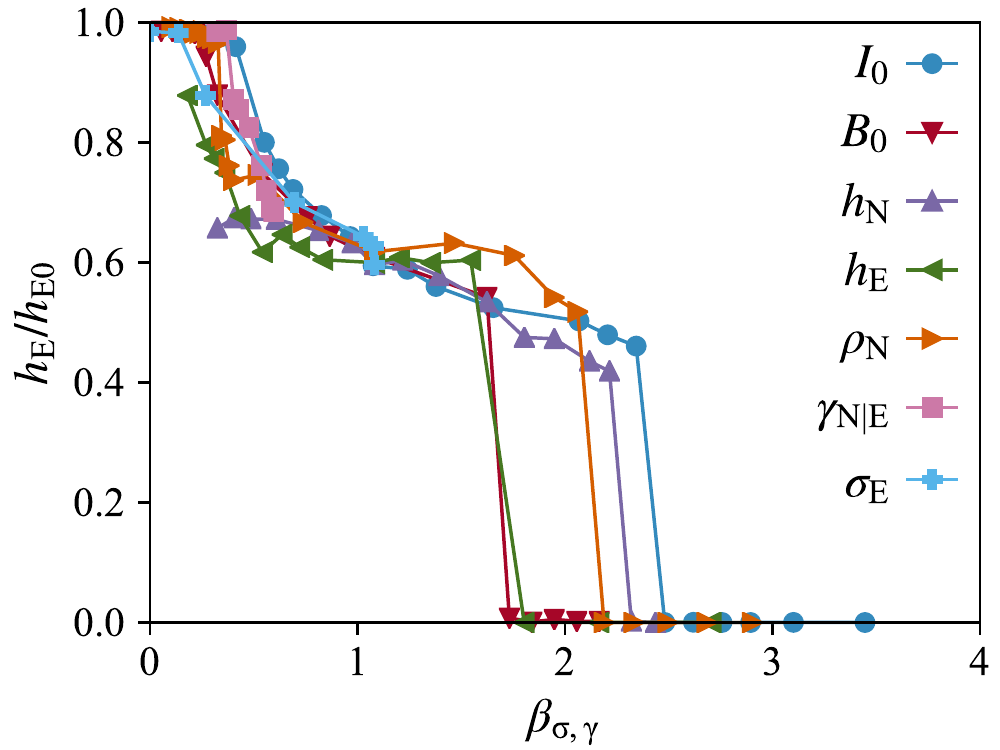}
    }
    \put(0.0,0.015){a)}
    \put(0.97,0.015){b)}
  \end{picture}
  \caption{Sketch of the current path $I_\text{c}$ and the
    compensation current $\Delta j$ in a cell with tilted interfaces
    (a). Modified ratio of minimal to undisturbed electrolyte height
    $h_\text{E}/h_{\text{E}0}$ vs.~modified Sele parameter
    $\beta_{\sigma, \gamma}$ according to
    Eq.~\eqref{eq:Sele_sigma_gamma} (b). Only the variable indicated in
    the legend is changed for each graph, all other values are kept
    constant.}
  \label{fig:Sele}
\end{figure}

Preventing electrolyte layer disruptions is the main motivation for
studying interfacial instabilities, the topic of the remainder of the
present paper. A detailed discussion of the various driving forces for
fluid motion in LMBs and consequences for their operation is provided
in \cite{KelleyWeier:2018}.

\section{Sloshing in liquid metal batteries}
The interfacial instability, which may occur in aluminum reduction
cells is commonly known as ``sloshing'' or ``metal pad roll
instability''. Sele \cite{Sele:1977} suggested to use the ratio of
Lorentz forces to gravity
\begin{equation}
  \beta_{\text{Sele}} = \frac{I_\text{c} B_0}{g \Delta \rho h_{\text{Al}}
    h_{\text{cryolite}}} > \beta_{\text{crit}} 
\label{eq:Sele}
\end{equation}
as a criterion for instability. $B_0$ denotes a vertical magnetic
field, $g$ gravity, $\Delta \rho$ the density difference between
aluminum and cryolite, and $h_{\text{Al}}$ and $h_{\text{cryolite}}$
the layer heights of aluminum and cryolite, respectively. Naturally,
only the horizontal components of the total cell current $I_\text{c}$ are
able to interact with the vertical magnetic field and
Eq.~\eqref{eq:Sele} takes this into account.

Depending on the
ratio of the density jumps at the two interfaces, Eq.~\eqref{eq:Sele}
can give good predictions for instabilities in LMBs as well
\cite{WeberBecksteinHerremanEtAl:2017}.  If the density jump at one
interface is much smaller than that on the other, as is the case for
the Mg$||$Sb LMB used in \cite{WeberBecksteinHerremanEtAl:2017}, the
system can be described in good approximation as having an oscillating
and a stationary interface. In this case, only the
Mg$|$Electrolyte (N$|$E) interface is markedly
deformed. Fig.~\ref{fig:Sele}b shows the ratio of the minimum
electrolyte layer height to its initial value
$h_\text{E}/h_{\text{E}0}$ vs.~a modified Sele parameter according to
Eq.~\eqref{eq:Sele_sigma_gamma}. This is Sele's equation
Eq.~\eqref{eq:Sele} supplemented by terms accounting for the conductivity
differences and the interfacial tension. It is the latter that allows
here for a better, but still imperfect collapse of the graphs
compared to Fig.~12 of \cite{WeberBecksteinHerremanEtAl:2017}.

A system with a stationary and a moving interface is enforced as well
in LMBs using a retainer for the negative electrode as sketched in
Fig.~\ref{fig:LMB_sketch}b. Compared to aluminium reduction cells,
tolerable electrolyte heights are limited to a few millimeters,
typically $h_{\text{E}}\le10$\,mm since the open circuit voltage
available for discharge does rarely exceed a value of 1\,V. 

If density jumps of comparable magnitude at both interfaces occur,
the system dynamic becomes richer as described in
\cite{HorstmannWeberWeier:2018}. However, not only the density ratios
of the layers have an influence on the hydrodynamics but also the
distribution of conductivities. This is the topic of the next section.

\section{Influence of the electrical conductivity distribution on sloshing}
The distribution and strength of the compensation currents $\Delta j$
sketched in Fig.~\ref{fig:Sele}a will depend on the conductivities of
the three layers N, E, and P. Most, if not all, stability criteria
developed for Hall-H{\'e}roult cells are rightly formulated without
taking conductivities into account because the material system is
quite fixed once and for all. This is not the case for LMBs, where
several different material combinations are under investigation.

Starting from the assumption that the current in the low conducting
electrolyte layer has to be a purely vertical one, the compensation current
$\Delta j$ in that layer can be modeled by starting from a series
connection with a total area specific resistance $R''_\text{total}$ given by
\begin{equation}
R''_\text{total} = R''_\text{N} + R''_\text{E} + R''_\text{P} =
\frac{h_\text{N}}{\sigma_\text{N}} +
\frac{h_\text{E}}{\sigma_\text{E}} + \frac{h_\text{P}}{\sigma_\text{P}} 
\label{eq:R_total}
\end{equation}
with $R''_\text{i}$, $h_\text{i}$, $\sigma_\text{i}$ denoting the area
specific resistance, height, and conductivity of the three layers N,
E, and P, respectively. By using the cell voltage to calculate the
difference between horizontal and inclined interfaces while allowing
only for small interface perturbations $\eta_{\text{N}|\text{E}}$,
$\eta_{\text{E}|\text{P}}$, one ends up with an expression for the
compensation current in the hydrodynamic two-layer but electromagnetic
three-layer case that is characteristic for the Mg$||$Sb cell:
\begin{equation}
  \Delta j_{\text{N}|\text{E}} = -\frac{I_0 (\sigma_\text{E} - \sigma_\text{N})\eta_{\text{N}|\text{E}}}{h_\text{N} \sigma_\text{E} + h_\text{E} \sigma_\text{N} + h_\text{P} \frac{\sigma_\text{N} \sigma_\text{E}}{\sigma_\text{P}}}.
  \label{eq:Delta_j_NE}
\end{equation}
From this perturbation current, a new Sele-like criterion can be formulated
by relating the Lorentz forces to the balancing forces. Following
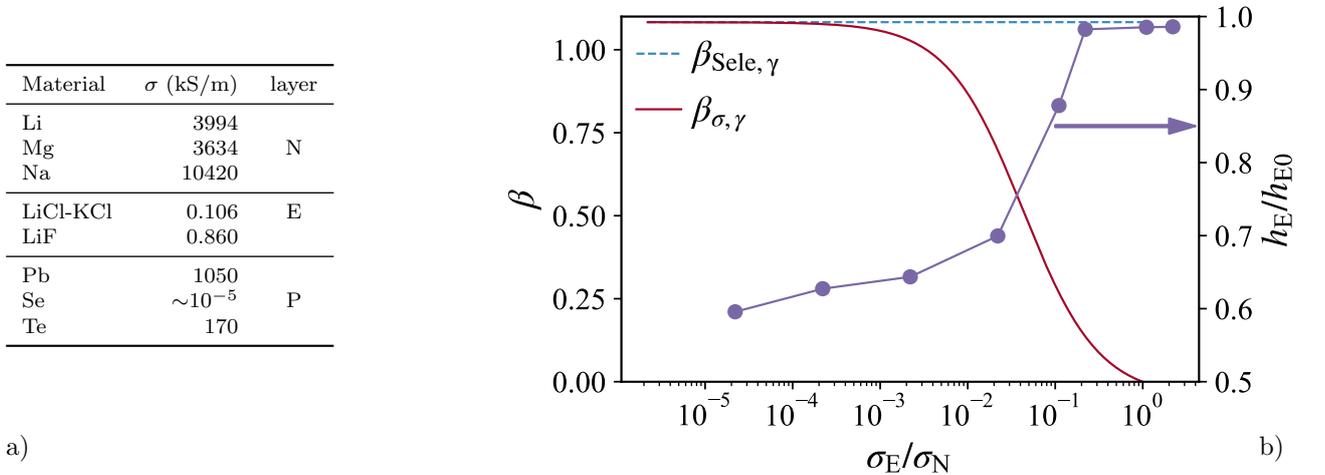
\begin{figure}[b!]
  \setlength{\unitlength}{\textwidth}
  \begin{picture}(1.0,0.36)
    \put(0.0,0.205){%
      \parbox[t]{0.3875\textwidth}{%
        \footnotesize
        \begin{tabular}{lrc}
          \toprule
          Material & $\sigma$ (kS/m) & layer \\
          \midrule
          Li & 3994 & \\
          Mg & 3634 & N \\
          Na & 10420 & \\
          \midrule
          LiCl-KCl & 0.106 & E \\
          LiF & 0.860 & \\
          \midrule
          Pb & 1050 &  \\
          Se & $\sim$10$^{-5}$ & P\\
          Te & 170 & \\
          \bottomrule
        \end{tabular}
      }
    }
    \put(0.3875,0.0){%
      \includegraphics[width=0.6125\textwidth]{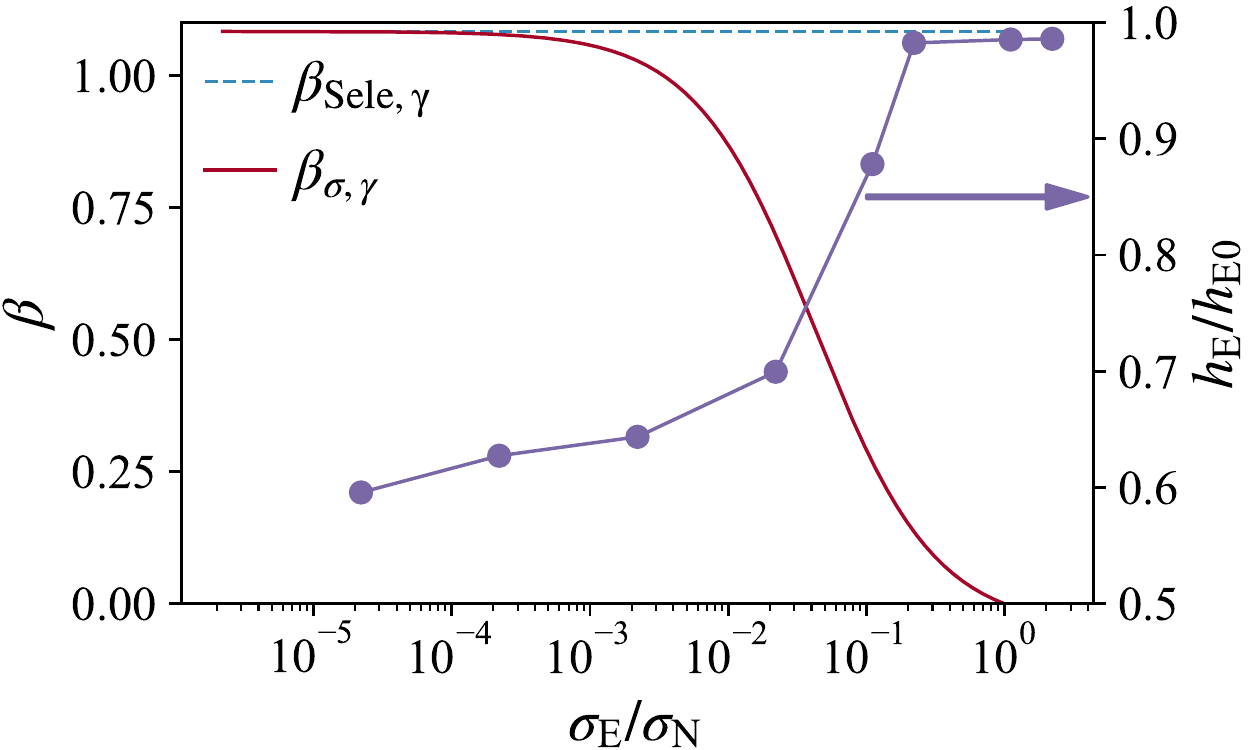}\\
    }
    \put(0.0,0.015){a)}
    \put(0.97,0.015){b)}
  \end{picture}
  \caption{Conductivity values of selected active and electrolyte
    materials at melting point \cite{KelleyWeier:2018,MassetHenryPoinsoPoignet:2006}, except Se
    $\sigma(480$\,\textdegree{}C) \cite{Lizell:1952} and Te
    $\sigma(452$\,\textdegree{}C) \cite{Perron:1967} (a). $\beta_{\text{Sele}, \gamma}$ and $\beta_{\sigma,
      \gamma}$ vs.~conductivity ratio and corresponding interface
    deformation (b).}
  \label{fig:conductivities}
\end{figure}
\cite{GerbeauLeLelievre:2006}, a Lorentz force
scaling due to \eqref{eq:Delta_j_NE} results as:
\begin{equation}
F_\text{L} = \frac{(\sigma_\text{N} -\sigma_\text{E})I_0 B_0
  \eta_{\text{N}|\text{E}}}{h_\text{N}(h_\text{N} \sigma_\text{E} +
  h_\text{E} \sigma_\text{N} + h_\text{P} \frac{\sigma_\text{N}
    \sigma_\text{E}}{\sigma_\text{P}})}.
\label{eq:F_Lorentz}
\end{equation}
Taking interfacial tension $\gamma_{\text{N}|\text{E}}$ into account, the balancing forces scale as
\begin{equation}
F_{g, \gamma} = [(\rho_\text{E} - \rho_\text{N})g +
\gamma_{\text{N}|\text{E}}\left(\frac{\epsilon_{mn}}{a}
\right)^2]\eta_{\text{N}|\text{E}},
\label{eq:F_gravity-capillary}
\end{equation}
cf.~\cite{HorstmannWeberWeier:2018}. $\epsilon_{mn}$ marks the
radial wavenumbers as described in \cite{HorstmannWeberWeier:2018},
while $a$ is the cell radius. In
many cases the $m=n=1$ mode, giving $\epsilon_{11} \approx 1.841$,
becomes most unstable to the metal pad roll instability. Taking the
ratio of both forces from Eqs.~\eqref{eq:F_Lorentz} and
\eqref{eq:F_gravity-capillary} leads to a new stability criterion
capturing the influence of the electrical conductivities as well as
the interfacial tension:
\begin{equation}
\beta_{\sigma, \gamma} = \frac{(\sigma_\text{N} -\sigma_\text{E})I_0
  B_0}{(h_\text{N} \sigma_\text{E} + h_\text{E} \sigma_\text{N} +
  h_\text{P} \frac{\sigma_\text{N}
    \sigma_\text{E}}{\sigma_\text{P}})h_\text{N} [(\rho_\text{E} -
  \rho_\text{N})g +
  \gamma_{\text{N}|\text{E}}\left(\frac{\epsilon_{mn}}{a} \right)^2]}.
\label{eq:Sele_sigma_gamma}
\end{equation}
Figure \ref{fig:conductivities}b shows $\beta_{\sigma, \gamma}$ and
the simplified version
\begin{equation}
\beta_{{\rm Sele}, \gamma} = \frac{I_0 B_0}{h_\text{N} h_\text{E}
  [(\rho_\text{E} - \rho_\text{N})g +
  \gamma_{\text{N}|\text{E}}\left(\frac{\epsilon_{mn}}{a} \right)^2]}
\label{eq:Sele_gamma}
\end{equation}
in dependence on $\sigma_\text{E} / \sigma_\text{N}$ using the
standard case parameters from
\cite{WeberBecksteinHerremanEtAl:2017}. For
$\sigma_\text{E} / \sigma_\text{N} \lesssim 3\cdot 10^{-4}$,
$\beta_{\sigma}$ is saturated and coincides with $\beta_{{\rm Sele}}$. In
that regime the assumption $\sigma_\text{E} \ll \sigma_\text{N}$ is
valid. In contrast, for
$\sigma_\text{E} / \sigma_\text{N} \gtrsim 3\cdot 10^{-4}$,
$\beta_{\sigma, \gamma}$ is rapidly decreasing, meaning that
destabilizing Lorentz forces decay and the system gets stabilized.  The
right ordinate of Fig.~\ref{fig:conductivities}b displays numerical
results for the minimal electrolyte layer height $h_\text{E}$ relative
to its initial value $h_\text{E0}$ in dependence on
$\sigma_\text{E} / \sigma_\text{N}$. These results show clearly that
the decrease of $\beta_{\sigma, \gamma}$ is indeed accompanied by a
decline of the interface deformations and therefore a stabilization of
the interface. Referring back, Fig.~\ref{fig:Sele}b contains data
points for variations of the interfacial tension
$\gamma_{\text{N}|\text{E}}$ and the conductivity $\sigma_\text{E}$
with all other parameters kept at constant values. The corresponding
data points fit quite well to the other curves supporting the
initial assumptions.

Looking at the conductivity values in Tab.~\ref{fig:conductivities}a,
it becomes apparent that the electrical conductivities of the negative
electrode and the electrolyte will probably fulfill the condition
$\sigma_\text{N} \gg \sigma_\text{E}$ in most cases, whereas the
analogous requirement $\sigma_\text{P} \gg \sigma_\text{E}$ is not
always met. However, the latter finding will most likely not be of
much practical relevance since both, tellurium as well as selenium,
are very scarce goods and unlikely to be used at scale for LMBs. In
addition, the conductivity of Se increases rapidly, if it is laced
with highly conducting impurities, as is the case, e.g., in a Li$||$Se
cell. Yet the above findings signal that one should take care when
artificially reducing the conductivity differences between active
materials and electrolyte in numerical LMB models
\cite{HerremanNoreCappaneraGuermond:2015}. In addition, it easily
explains why model experiments using metal combinations with
miscibility gaps (e.g., Li$|$Na, Ga$|$Pb, Bi$|$Ga, see table
\ref{tab:miscibility_gap} for a selection) to study
electromagnetically excited interfacial instabilities will face
difficulties in achieving measurable interface deformations with
moderate currents and magnetic fields.

\section{Liquid metal two-layer systems}
To quantifiy interface instabilities in liquid metal two layer
\begin{table}
  \caption{\label{tab:miscibility_gap}Selection of binary alloy
    systems possessing a miscibility gap and approximate monotectic
    ($T_\text{m}$, melting temperature zinc in case of Na-Zn) and
    critical ($T_\text{K}$) temperatures. The metals in the rows
    `upper layer' and `lower layer' are the main components of the
    alloys near $T_\text{m}$ and $\rho_\text{upper}$,
    $\rho_\text{lower}$ the densities of the pure components at
    $T_\text{m}$ (various sources). $P$ and $\sigma_\text{upper}/\sigma_\text{lower}$
    denote the ultrasound reflectance and the ratio of
    conductivities, respectively, of the pure components at
    $T_\text{m}$. The higher $P$, the stronger is the reflection of an
    ultrasound beam incidenting perpendicularly to the interface.}
  \centering
\begin{tabular}{llllrrcc}
  \hline
  lower & upper & $T_\text{m}$ & $T_\text{K}$ & $\rho_\text{lower}$ &
                                                                      $\rho_\text{upper}$
  & $P$ & \rule{0pt}{1.1em}  $\frac{\sigma_\text{upper}}{\sigma_\text{lower}}$\\
layer & layer & $\mbox{}^\circ$C & $\mbox{}^\circ$C & kg m$^{-3}$ & kg
                                                                    m$^{-3}$
  & & \\ \hline
   Bi & Ga & 222 & 258 &  10108 &   5992 & 0.00014 & 4.31 \\
   Bi & Zn & 416 & 580 &   9879 &   6583 & 0.00492 & 3.69 \\
   Hg & Ga & 27  & 204 &  13531 &   6102 & 0.00185 & 3.74 \\
   Pb & Ga & 313 & 609 &  10689 &   5941 & 0.00415 & 3.02 \\
   Na & Li & 171 & 303 &    911 &    515 & 0.00030 & 0.50 \\
   Zn & Na & 420 & 819 &   6579 &    851 & 0.64974 & 1.64 \\
   Pb & Zn & 420 & 802 &  10548 &   6579 & 0.00000 & 2.66 \\
BiPb$_\text{e}$ & Ga & 113 & 514 &  10566 &   6053 & 0.00078 & 3.99 \\ \hline
  \end{tabular}
\end{table}
systems, we discuss a few exemplary cases in the following. The
forementioned difficulties notwithstanding, liquid metal systems
possessing a miscibility gap have the advantage of easily supporting
high current densities without associated chemical reactions since
electrons are the charge carriers in both layers. These liquid metal
systems might therefore be candidates for model experiments to study
electromagnetically induced interface deformations. Besides the ratio
of electrical conductivities, the densities and the acoustic
properties of the pure components are listed in Tab.~\ref{tab:miscibility_gap}. Two immiscible liquids exist above the
monotectic temperature (T$_\text{m}$, melting temperature zinc in case
of Na-Zn) and below the critical temperature T$_\text{K}$, where the
miscibility gap closes. An exemplary 
\begin{figure}
  \setlength{\unitlength}{\textwidth}
  \begin{picture}(1.0,0.47)
    \put(0.0,0.0){%
      \includegraphics[width=0.67\textwidth]{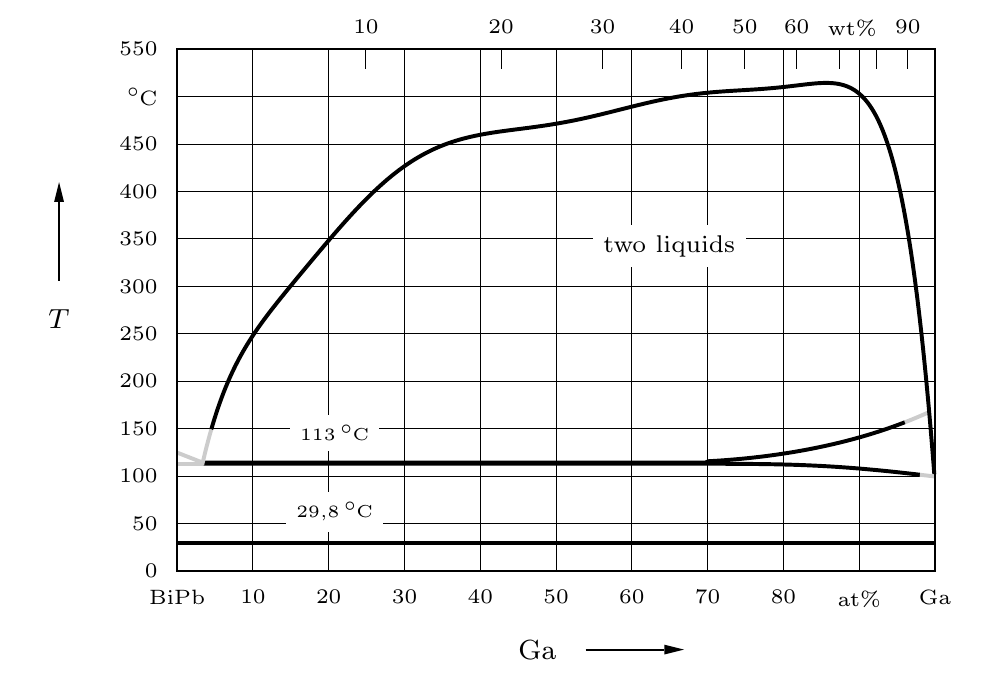}
    }
    \put(0.65,0.15){%
      \includegraphics[width=0.32\textwidth]{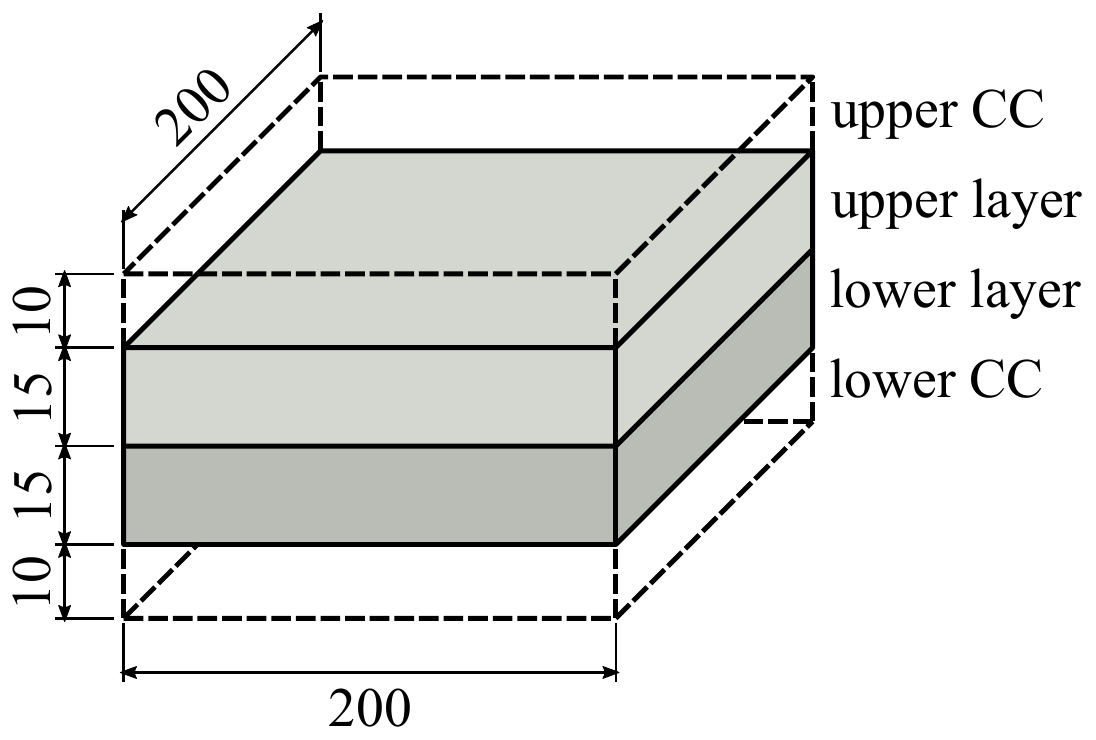}
    }
    \put(0.0,0.01){a)}
    \put(0.67,0.12){b)}
    \end{picture}
    \caption{a) Experimental isopleth in the Bi-Ga-Pb system at
      $x_\text{Pb}= 0.808 x_\text{Bi}$ determined by simple cooling
      curves. b) sketch (not to scale, measures in mm) of the two layer
      arrangement. The depicted current collectors
      (CCs) were considered for a subset of the discussed cases only.}
\label{fig:isopleth}
\end{figure}
system with a miscibility gab is shown in
Fig.~\ref{fig:isopleth}a. The system consists of an eutectic Bi-Pb
alloy (55.3-44.7\,mol\%) as the lower and Ga as the upper layer and is
therefore a ternary one. The depicted isopleth is for the eutectic
Bi-Pb composition, i.e. $x_\text{Pb}= 0.808 x_\text{Bi}$. As a result,
there is no well defined monotectic temperature at the Ga rich side,
but probably a narrow mixed phase region. Detailed information could
not be obtained by the simple cooling curve technique applied to
generate the phase diagram. Besides this, the miscibility gap forms at
relatively low temperature and persists for about 400\,K. Interface
detection by ultrasound would, however, be difficult because of the
low reflectance of the interface. This property is shared by most of
the listed systems except Na$|$Zn. The conductivity ratio of pure Ga
to eutectic Bi-Pb is with almost 4 comparably high. It is outmatched
only by Bi$|$Ga with 4.31. Experimentally, a conductivity ratio of
about 2.5 can be measured at around 230\,$\mbox{}^{\circ}$C for
Bi$|$Ga, because the conductivity contrast of the Bi-rich and Ga-rich
phases existing in reality is lower than that of the pure components,
especially for the relatively flat miscibility gap of the Bi-Ga
system.

Excitation of interfacial waves by an externally applied vertical
current in the presence of a vertical magnetic field was studied
numerically with an OpenFOAM based solver using the volume of fluid
method. Material properties of Bi-Ga at 222\,$\mbox{}^{\circ}$C are
used for the simulations. Figure \ref{fig:isopleth}b displays a sketch
of the geometry. Both liquid layers are 15\,mm in height and have a
cross section of 200$\times$200\,mm$^2$. Current is applied to the top
and the bottom surfaces. The setup is modeled with four different
approximations: 1) Dirichlet boundary conditions directly at the
liquids' surfaces (constant electric potential, DL), 2) Neumann
boundary conditions directly at the liquids' surfaces (constant
current, NL), 3) Dirichlet boundary conditions at the current
collectors (DCC), and 4) Neumann boundary conditions at the current
collectors (NCC).
\begin{figure}
  \setlength{\unitlength}{\textwidth}
  \begin{picture}(1.0,0.46)
    \put(0.0,0.012){%
      \includegraphics[width=0.48\textwidth]{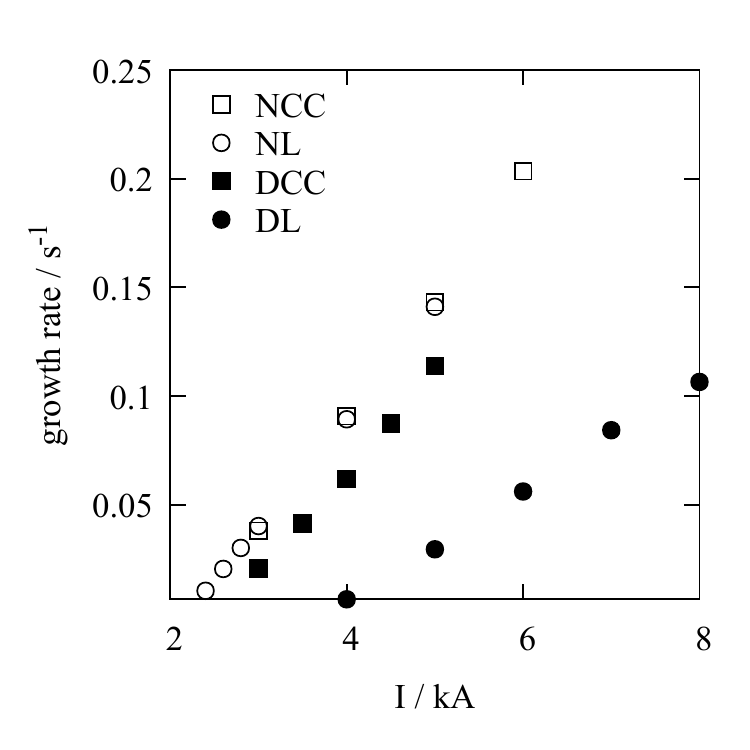}
    }
    \put(0.00,0.04){a)}
    \put(0.52,0.04){b)}
    \put(0.52,0.0){%
      \includegraphics[width=0.48\textwidth]{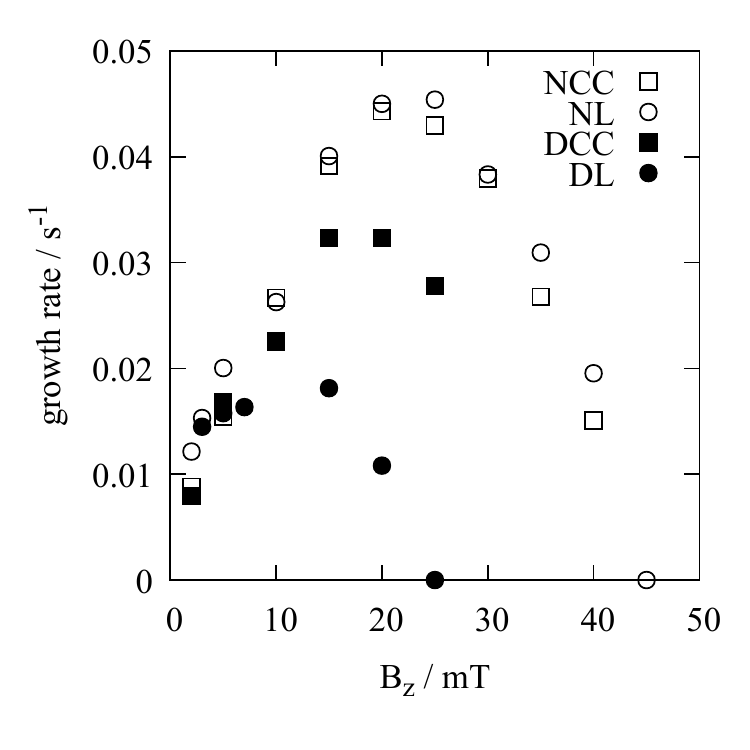}
    }
\end{picture}
\caption{Growth rates of the interfacial waves for a constant
  $B_z=30$\,mT vs.~current (a) and for constant $I=3$\,kA vs.~$B_z$
  (b). NCC\dots Neumann b.c.~at the current collectors, NL\dots
  Neumann b.c.~at the liquids' surfaces, DCC\dots Dirichlet b.c.~at
  the current collectors, DL\dots Dirichlet b.c.~at the liquids'
  surfaces.}
  \label{fig:growth_rates}
\end{figure}

In the first case (DL, Dirichlet b.c.~at the liquids' surfaces), the
first growing interfacial waves are detected for a critical current of
$I_\text{cr}=3.78$\,kA at an already quite strong magnetic field of
$B_z=30$\,mT. As can be seen in Fig.~\ref{fig:growth_rates}a), the
growth rates increase linearly for
$I>I_\text{cr}$. Fig.~\ref{fig:growth_rates}b) displays the growth
rates of the waves for a constant current $I=3$\,kA vs.~$B_z$. This
plot reveals that Fig.~\ref{fig:growth_rates}a) does not show the most
critical instability. Instead $B_z=30$\,mT is so high that it already
damps waves that would occur for lower $B_z$ and lower $I$. This
behavior was recently found by Herreman et al.~\cite{HerremanNoreGuermondCappaneraWeberHorstmann:2019} for
cylindrical reduction cells which give a more complete picture including
stability curves in the $I$-$B_z$-plane (Fig.~5 of
\cite{HerremanNoreGuermondCappaneraWeberHorstmann:2019}). The two cuts
through this plane represented by Fig.~\ref{fig:growth_rates}a) and
Fig.~\ref{fig:growth_rates}b) demonstrate that cells with square cross
sections basically behave in a similar manner. Due to the high
conductivities of both layers, the influence of magnetic damping on
the waves is
large
and explains the decrease of the growth rates for $B_z > 15$\,mT.

Looking at the growth rates obtained with Neumann boundary conditions
at the liquids' surfaces (second case, NL), a considerably lower
critical current of $I_\text{cr}=2.2$\,kA is observed and again a
linear increase of the growth rate with the current. Correspondingly,
wave damping in the NL case starts to decrease the growth rates later
($B_z > 25$\,mT) and completely cancels them only at about
$B_z = 45$\,mT. Neumann b.c.~force the perturbation currents due to
the interface inclination to close in the liquid layers, while
Dirichlet b.c.~do not impose such restrictions. Consequently, wave
exciting Lorentz forces are stronger under Neumann b.c., resulting in
lower critical currents, larger growth rates, and deferred
overdamping.

Taking the current collectors (modeled with the conductivity of
stainless steel that is slightly lower than that of the liquid metals)
into account, changes especially the wave evolution for Dirichlet
b.c. (DCC). The critical current for $B_z = 30$\,mT reduces to
$I_\text{cr}=2.6$\,kA and the growth rate increases almost in parallel
with the Neumann b.c. cases. In Fig.~\ref{fig:growth_rates}b) damping
decreases growth rates now for larger magnetic fields ($B_z > 20$\,mT)
compared to DL. This might be explained again by recurring to the
horizontal perturbation currents. The lower conductivity of the
current collectors effectively forces a larger part of the
perturbation currents to close in the liquid metals. As mentioned
before, Neumann b.c. maximize this effect, hence adding the current
collectors to this case (NCC) does not markedly change the growth
rates compared to the NL case.

\section{Conclusions}
LMBs are electricity storage devices with a completely or dominantly
liquid interior. Accordingly, fluid dynamics and magnetohydrodynamics
are helpful in understanding and improving the operation
characteristics of LMBs. Fluid flow can be generated by a number of
mechanisms and may have different consequences. On the one hand, those
can be beneficial as in the improved mixing inside the positive
electrode that will foster capacity usage and prevent the growth of
intermetallics. On the other hand, larger deformations of the
interfaces should be avoided in order to keep the delicate electrolyte
layer intact and thereby to avoid short-circuits.

Electromagnetically excited gravity waves may arise in the presence of
a vertical magnetic field due to slight inclinations of the
interface. The horizontal perturbation currents thereby evoked depend
on the conductivities of the single layers. The lower the conductivity
differences, the higher the critical current for wave amplification. A
modified Sele parameter incorporating the conductivity ratios was
proposed to take this effect into account.

Two layer liquid metal systems possessing a miscibility gap can be
used to study interfacial wave dynamics in the lab. However, one has
to resort to high electric currents and to cope with substantial
magnetic damping. Boundary conditions have a marked influence on
computed growth rates and the extension of the unstable region in the
$I$-$B_z$-plane.

\section*{Acknowledgement}
  This work was supported by the Deutsche Forschungsgemeinschaft
  (DFG, German Research Foundation) by award number 338560565, by the
  Latvian Council of Sciences under project number lzp-2018/1-0017,
  and in frame of of the Helmholtz - RSF Joint Research Group
  ``Magnetohydrodynamic instabilities: Crucial relevance for large
  scale liquid metal batteries and the sun-climate connection'',
  contract No HRSF-0044 and RSF-18-41-06201. The computations were
  performed on the Bull HPC-Cluster “Taurus” at the Center for
  Information Services and High Performance Computing (ZIH) at TU
  Dresden and on the cluster “Hydra” at Helmholtz-Zentrum Dresden –
  Rossendorf. Fruitful discussions with Valdis Bojarevics, Wietze
  Herreman, Douglas Kelley, Caroline Nore, Donald Sadoway, and Oleg
  Zikanov on several aspects of metal pad roll instabilities and
  liquid metal batteries are gratefully acknowledged. N.~Weber thanks
  Henrik Schulz for the HPC support.

\bibliographystyle{unsrt}
\bibliography{bibexport}

\end{document}